# Carrier thermalization dynamics in single Zincblende and Wurtzite InP nanowires

*Yuda Wang,   Howard E. Jackson   and   Leigh M. Smith*

Department of Physics, University of Cincinnati, Cincinnati, OH 45221-0011

*Tim Burgess,   Suriati Paiman,   Qiang Gao,   Hark Hoe Tan   and   Chennupati Jagadish*

Department of Electronic Materials Engineering, Research School of Physics and Engineering,

The Australian National University, Canberra, ACT 0200, Australia

e-mail address:  leigh.smith@uc.edu

ABSTRACT

Using transient Rayleigh scattering (TRS) measurements, we obtain photoexcited carrier thermalization dynamics for both zincblende (ZB) and wurtzite (WZ) InP single nanowires (NW) with picosecond resolution. A phenomenological fitting model based on direct band to band transition theory is developed to extract the electron-hole-plasma density and temperature as a function of time from TRS measurements of single nanowires which have complex valence band structures. We find that the thermalization dynamics of hot carriers depends strongly on material (GaAs NW vs. InP NW) and less strongly on crystal structure (ZB vs. WZ). The thermalization dynamics of ZB and WZ InP NWs are similar. But a comparison of the thermalization dynamics in ZB and WZ InP NWs with ZB GaAs NW reveals more than an order of magnitude slower relaxation for the InP NWs. We interpret these results as reflecting their distinctive phonon band structures which lead to different hot phonon effects. Knowledge of hot carrier thermalization dynamics is an essential component for effective incorporation of nanowire materials into electronic devices.

# Introduction

Recent dramatic advances in the growth of semiconductor nanowire (NW) and its heterostructures have been followed by substantial development of novel devices utilizing these structures [1] including nano-electronics [2-3], nano-photonics [4-6], nano-biochemistry [7-9] and nano-energy [10-11] applications. The design of these applications depends critically on knowledge of the relaxation of hot carriers in the semiconductor NWs. Previously, Montazeri et al. [12-13] used time-resolved Rayleigh scattering to obtain detailed information on carrier thermalization in GaAs NWs. How these relaxations depend on the NW material, e.g. GaAs or InP, or crystal structure, e.g. ZB or WZ, is an open question. Here, using a simple analysis of detailed TRS spectra in both ZB and WZ InP NWs, we find that the dynamics in both these crystal structures are dramatically different from what has been observed in ZB GaAs NWs.

While in nature InP exists only in the ZB phase, in nanowire structurers both ZB and WZ polytypes can be fabricated with high crystal and optical quality, and homostructures using both polytypes can be made. While the electronic band structures of ZB and WZ InP nanowires are now well understood [14-17], there is no information on the thermalization of carriers in WZ InP, and extremely limited information for ZB InP, even in bulk material [18-21].

In this paper, for the first time, we provide a complete picture of the carrier cooling dynamics of both ZB and WZ InP in picosecond to nanosecond regime at 10K low temperature with pump-probe transient Rayleigh scattering (TRS) measurements by expanding the analysis of Montazeri et al. [12-13] to include complex multi-band structures. For instance, in WZ InP NWs, relaxation dynamics must include both electron thermalization in a single conduction band, as well as holes in three separate valence bands: the A (heavy hole), B (light hole) and C bands [15-16, 23] which are separated by energies determined by the spin-orbit and crystal-field interactions [22]. To interpret the experimental results, we develop a simple formalism based on direct band-band theory which can model the changes in the real and imaginary parts of the index of refraction when the bands are dynamically occupied by carriers. The many-body effects are included phenomenologically. We extract the electron-hole density and temperature as a function of time for each polytype for all bands, as well as their energy positions. We then provide a phenomenological explanation of the contrasting dynamics of InP and GaAs NWs by considering hot phonon effects

## TRS experimental results and spectra fitting model

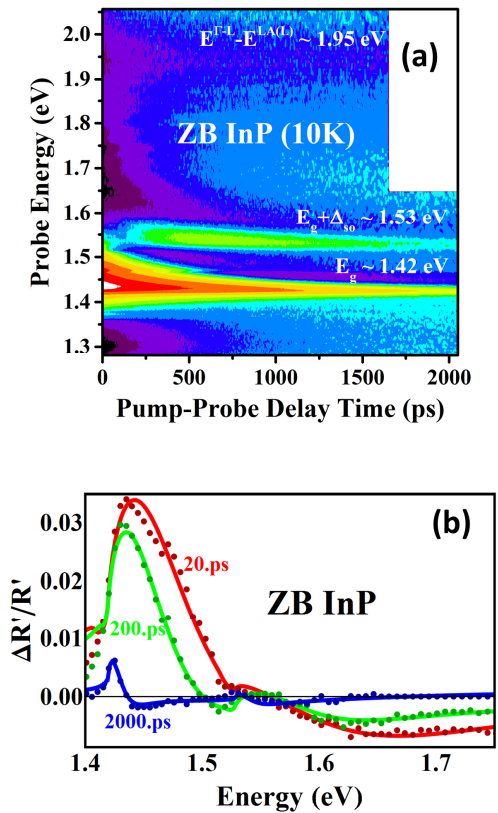

**Figure 1**. ZB InP NW TRS results and fits: (**a**) False color map of ΔR'/R' data, as a function of pump-probe delay time and probe energy. (red – positive, blue – negative). The band edge

transitions can be resolved at late times (~ 2 ns) and are consistent with measurements by other methods. The early time signal is broadened and shifted towards higher energy primarily due to band filling effects. The signal decay evolution is clear when one moves from early to late times. (**b**) Selected ΔR'/R' vs energy spectra (dots are measurements) for times after pump pulse of 20 ps (red), 200 ps (green) and 2000 ps (blue) as well as fitting (solid) curves based on direct band to band transition theory. The EHP density and temperature in all bands are revealed through this analysis (e.g. at 20 / 200 / 2000 ps after initial excitation, the electron densities are 1 × $10^{18}$ / 8 × $10^{17}$ / 8 × $10^{16}$ cm$^{-3}$ and temperatures are 384 / 252 / 60 K, respectively).

These measurements are performed on single MOCVD grown ZB or WZ InP NWs at 10 K. The single phase ZB or WZ InP NWs are nominally 150 nm in diameter. The ZB NWs were intentionally doped by a low Si-doping concentration of approximately $10^{16}$ cm$^{-3}$ (deduced from TRS data analysis), while the WZ NWs were undoped. In these TRS experiments, the polarization of a tunable probe pulse (1.2 - 2.0 eV) oscillates between parallel and perpendicular to a single nanowire at 100 kHz using a photoelastic modulator. Using a balanced detector the back-scattered probe pulse is monitored by a lockin amplifier tuned to the 100 kHz PEM oscillator which provides the signal which is the polarized scattering efficiency: R'=$R_{\parallel}$-$R_{\perp}$. Using a pump pulse (at 2.25 eV) carriers are photoexcited into the wire which changes the real and imaginary part of the index of refraction which in turn changes the polarized scattering efficiency, R'. Both pump and probe laser pulses have a width of ~ 200 fs. The pump is modulated on and off by a mechanical chopper at 773 Hz. By looking at the change of R' when pump is on and off, we acquired ΔR' which is a derivative-like structure and is much sharper than R'. Moreover, the ΔR' is measured at different time delays of the probe after the initial pump by adjusting the length of the time delay line in the probe laser path. With such an experimental setup, we can measure the relative change of the polarized Rayleigh scattering efficiency ΔR'/R' as a function of both probe energy and probe delay time after the pump.

A typical spectrum from a single ZB InP NW is shown in Fig. 1b, and the spectra map was plotted as a function of time in Fig. 1a. The false color in the TRS map represents the value of ΔR'/R', with red showing positive and blue showing negative changes in the scattering efficiency R', where R' is the linear dichroism ($R_{\parallel}-R_{\perp}$). The derivative-like structure becomes very sharp at late times when the carrier occupation densities and temperatures are low, and so the energy position of the fundamental band to band transition is seen with high accuracy. The zero crossing point of the TRS spectrum is a direct measure of the energy associated with a direct band to band transition. The ZB InP NW band gap $E_g$ is observed at 1.42 eV corresponding to an optical transition between the degenerate light-hole (LH) & heavy hole (HH) bands and the conduction band, while the corresponding transition between the split-off (SO) band to the conduction band is observed at 110 meV higher energy. Both of the results are in excellent agreement with previous reports [23-26]. The ΔR'/R' spectral map (Fig.1a) at early times displays a more intense response (larger changes in R'), exhibit an increased energy width and is centered at higher energy compared with late times. At very early times (t < 100 ps), the broadening effects are strong enough to cause overlapping responses between the higher lying SO-e transition and the lower lying LH&HH-e band edge. Perhaps more surprisingly, a very weak transition is also observed around 1.95 eV which may be associated with the Γ-L transition between the top of the valence band and the conduction band L-valley.

To model these light scattering lineshapes, we calculate ΔR'/R' based on the direct band to band transitions theory where the bands are occupied by hot electrons and holes. The absorption coefficient $\alpha[E,N,T]$ (Fig. 2a) is calculated as a function of carrier density and temperature through [27]

$$\alpha[E,N,T] = \frac{\pi^2 c^2 h^3}{n^2 E^2 (2\pi)^3} B \int_0^{E-E_g} \rho_C[E'] \times \rho_V[E'-E] \times (f_l[E-E_g-E'] - f_u[E'])dE',$$

where n is the average index of refraction; B is the radiative bimolecular coefficient which depends on the transition matrix element; $\rho_i[E]$ is the 3D density of states in the conduction/valence band; $f_i[E] = \frac{1}{1+Exp[\frac{E-E_F}{k_B T}]}$ is the Fermi-Dirac distribution probability that upper/lower states involved in the transition are occupied by electrons, with the quasi-Fermi-energy $E_F[N,T]$ related to both the carrier density N and temperature T; $E'$ is the upper state energy above the conduction band minimum. For n and B, we use the average values from the literature [28-35] since only the relative change of the absorption coefficient is critical, not its absolute value. Using the Kramers-Kronig relation, we transform the calculated absorption coefficient $\alpha[E,N,T]$ to obtain the index of refraction $n[E,N,T]$ as a function of carrier density and temperature. We define the complex refractive index $\boldsymbol{n} = n + i\,k$, where the extinction coefficient $k = \frac{\lambda}{4\pi}\alpha$. As shown by Montazeri et al. [13], the lineshape ΔR'/R' is the direct product of a complex phase factor $e^{i\theta}$ containing NW geometrical information and the change of the complex index of refraction $\Delta\boldsymbol{n} = \Delta n + i\,\Delta k$. We therefore can write

$\frac{\Delta R'}{R'} = A\,e^{i\theta}\left(\Delta n + i\Delta\alpha\frac{\lambda}{4\pi}\right) = A\left\{cos[\theta]\,\Delta n - \frac{\lambda}{4\pi}sin[\theta]\,\Delta\alpha\right\}$, where A is an overall arbitrary amplitude factor. In Fig.2 we show the resulting coefficients α, n, Δα, Δn and ΔR'/R' for the HH-e optical transition corresponding to fits of the ZB InP TRS spectra at various times.

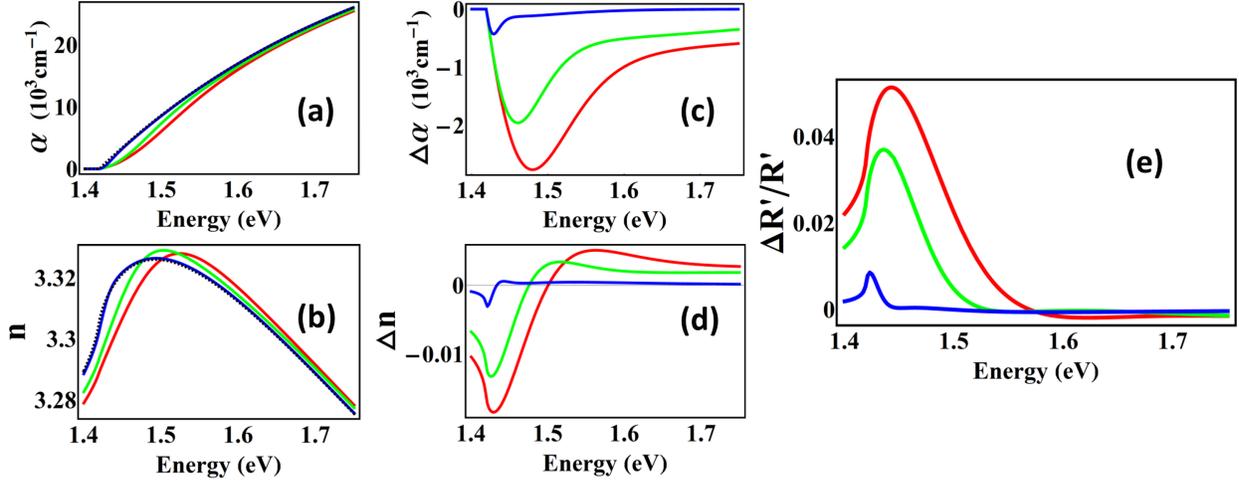

**Figure 2.** (**a**) and (**b**) Solid lines show the calculated absorption coefficient α[E] and index of refraction n[E] as a function of energy at different times (20, 200 and 2000 ps shown as red, green and blue respectively) after initial pump-excitation obtained from the spectral lineshape fitting displayed in Fig. 1b. The background $α_o[E]$ and $n_o[E]$ (Dashed) are calculated with no pump-photo-excitation but only probe-excitation (i.e. log $N_o = 10^{15}$ cm$^{-3}$, $T_o$=10 K). (**c**) & (**d**) show the *change* of the absorption coefficient Δα[E] and index of refraction Δn[E] at different times after initial pump with respect to background $α_o$ and $n_o$. (**e**) The final Rayleigh scattering efficiency $\frac{\Delta R'}{R'} \propto \left(\Delta n + i\Delta\alpha \frac{\lambda}{4\pi}\right)e^{i\theta}$ is represented as a direct product of the complex index of refraction and complex phase factor related to nanowire geometry.

The final fitting curve is simply the superposition of each ΔR'/R' curve for transitions from each of the three valence bands to the conduction band using the following assumptions: (1) The temperature of the electrons and holes are the same [36-38] (i.e. they are thermalized with each other through rapid (< 1 ps) carrier-carrier scattering after initial excitation.); (2) The sum of the hole densities in all 3 valence bands (HH, LH, SO for ZB / A, B, C for WZ) equals the density of electrons in the conduction band (i.e. charge neutrality is assumed); (3) The background density without pump excitation is $10^{15}$ cm$^{-3}$ (i.e. the probe pulse itself excites a few carriers), which means that separate quasi Fermi energies are defined for each band even if we assume the pump-excited holes are in thermal equilibrium between various hole bands; (4) The background temperature for these measurements is 10 K; (5) The amplitude factor A and bimolecular coefficient B are the same if similar electronic selection rules [39-40] exist among different transitions (The HH-e transition intensity dominates over LH-e or SO-e in ZB InP, because the HH mass dominates [23, 41] where density of states effective mass is one of the prefactors of n and α, which turn out to be the component of the modeled ΔR'/R' ); (6) Band-gap-renormalization (BGR) effects [42-43] occur due to the many-body-effects of the EHP and are treated as a free time-dependent fitting parameter causing a rigid shift in the band gap.

To fit the entire TRS spectra as a function of time, we optimize all the parameters by running iteration cycles. Within each cycle, we keep time-independent parameters fixed but tune and optimize time-dependent parameters to acquire the best fits for all the spectra. Those time-dependent parameters are carrier densities, common carrier temperatures and the BGR coefficients. The time-independent parameters are the band gap, SO split-off energy, amplitude factors, NW geometry phase factor as well as the background density and temperature. After one cycle is finished, we adjust time-independent parameters slightly and start another iteration cycle. By continually comparing the fitting results from different iterations, we finally acquire the optimized time-independent parameters as well as the corresponding time-dependent parameters. The results of these fits are selectively shown at several times after the pump pulse in Fig. 1b. The fitted band gap for ZB InP NW is at $1.419 \pm 0.001$ eV and the split-off band is at $0.108 \pm 0.001$ eV below the top of the valence band. These transition energies agree well with previous reports [23-26] and direct observations of the derivative structures at late times (Fig.1a). The fitted phase factor result indicates the diameter of the NW is approximately 160nm based on the relation between phase factor and nanowire diameter [12]. In analyzing the fits to the $\Delta R'/R'$ spectrum, we find that the intensity is primarily related to the hot carrier concentration while the spectral width is primarily related to the hot carrier temperature and band filling. As a result of these effects, the scattering lineshapes can overlap strongly at early times when both the density and temperatures are high. This overlap shows up, for instance, for the ZB InP $E_g + \Delta_{so}$ transition near 100 ps where both the intensity and energy appear to drop. This structure results from the superposition of the spectral lineshapes at $E_g$ and $E_g + \Delta_{so}$. At the earliest times, the band gap renormalization tend to reduce the band gap slightly but the strong band filling effects eventually cause the $\Delta R'/R'$ structure move to higher energy. Due to the small electron effective mass compared to holes, this band filling effect is dominated by the pump-modulated electron distribution in the conduction band.

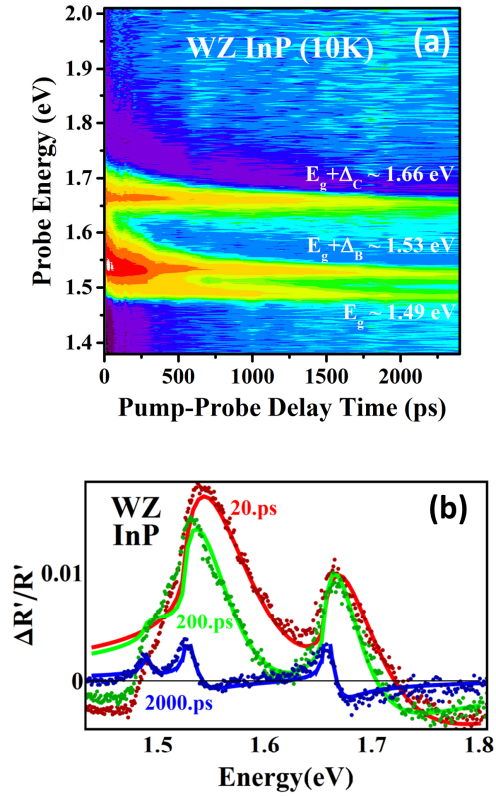

**Figure 3.** WZ InP nanowire TRS results and fits: (**a**) False color map of ΔR'/R' data as a function of pump-probe delay time and probe energy. (red – positive, blue – negative) (**b**) Selected ΔR'/R' spectra (dots are measurements) as well as fitted (solid) curves based on direct band to band transition theory.

Similar TRS measurements have been carried out on a single MOCVD grown 150 nm diameter undoped WZ InP NW at pump excitation intensity $4 \times 10^5$ kW cm$^{-2}$, as displayed in Fig. 3a. By looking at the zero point of the derivative-like structure at late times, the WZ InP NW $E_g$ is observed at ~1.49 eV which is the energy of the optical transition between the A valence band and the conduction band, while the B valence band to conduction band transition is observed ~30 meV higher in energy and the C valence band to conduction band transition is seen ~160 meV higher in energy. At a larger pump excitation intensity $2 \times 10^6$ kW cm$^{-2}$, the A and B valence band to second conduction band can be resolved at 1.74 eV and 1.77 eV (see Fig. s1). These results are in excellent agreement with recent reports [15-16, 22, 44]. Similar band filling behavior to that observed in ZB InP is also seen in the WZ InP NWs. We use the same fitting model based on direct band to band transition theory to extract the WZ InP NW EHP density and temperature as a function of time. One difference for the WZ InP TRS data fitting model is that the A-e transition intensity is extremely weak compared with B-e or C-e. The reason for this is that the A-e transition is only allowed for an electric-field polarized perpendicular to the WZ InP c-axis (NW long-axis), but this field is suppressed by the dielectric contrast between NW (150 nm diameter much smaller than laser wavelength ~800 nm) and the surrounding media (vacuum). The results of these fits are shown at several times after the pump pulse in Fig. 3b. The fitted A, B, C-valence band to conduction band energies for WZ InP are $1.485 \pm 0.005$ eV, $1.523 \pm 0.005$ eV and $1.654 \pm 0.005$ eV respectively, which agree well with previous reports [15-16] and direct observations of the derivative structures at late times (Fig. 3a). The fitted phase factor result gives an approximate NW diameter of 160 nm based on the relation between phase factor and nanowire diameter [12]; the actual diameter measured by TEM is about 150 nm [22].

## EHP density & temperature relaxation

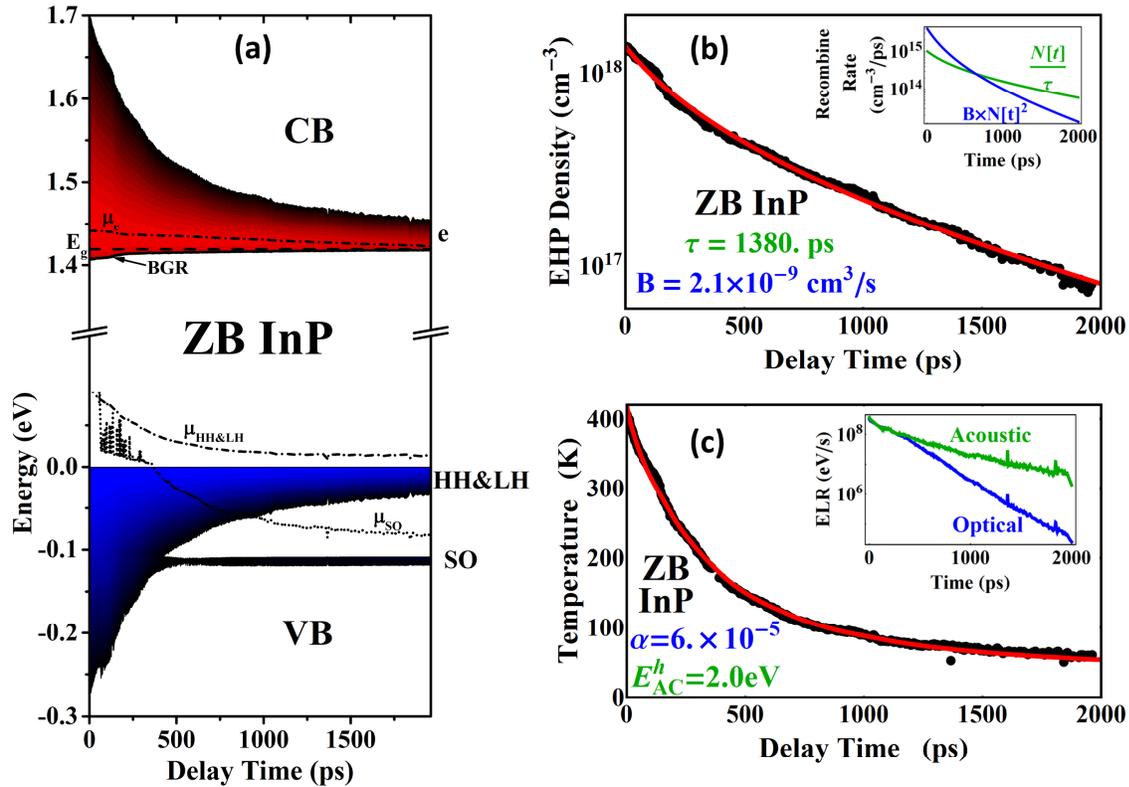

Fig.4 ZB InP nanowire TRS data fit results: (**a**) Electron-hole population density distributions, including quasi-Fermi-energies (dashed & dotted) and BGR (solid) in both conduction and valence bands as a function of time. (**b**) EHP density fit results (dots) as a function of time, with the red curve modeled by a sum of linear and bimolecular recombination processes. The inset shows the separate recombination rates. (**c**) EHP temperature fit results (dots) as a function of time, with the red curve modeled by optical and acoustic phonon scattering mechanisms. The inset shows the energy loss rates due to optical or acoustic phonon emission.

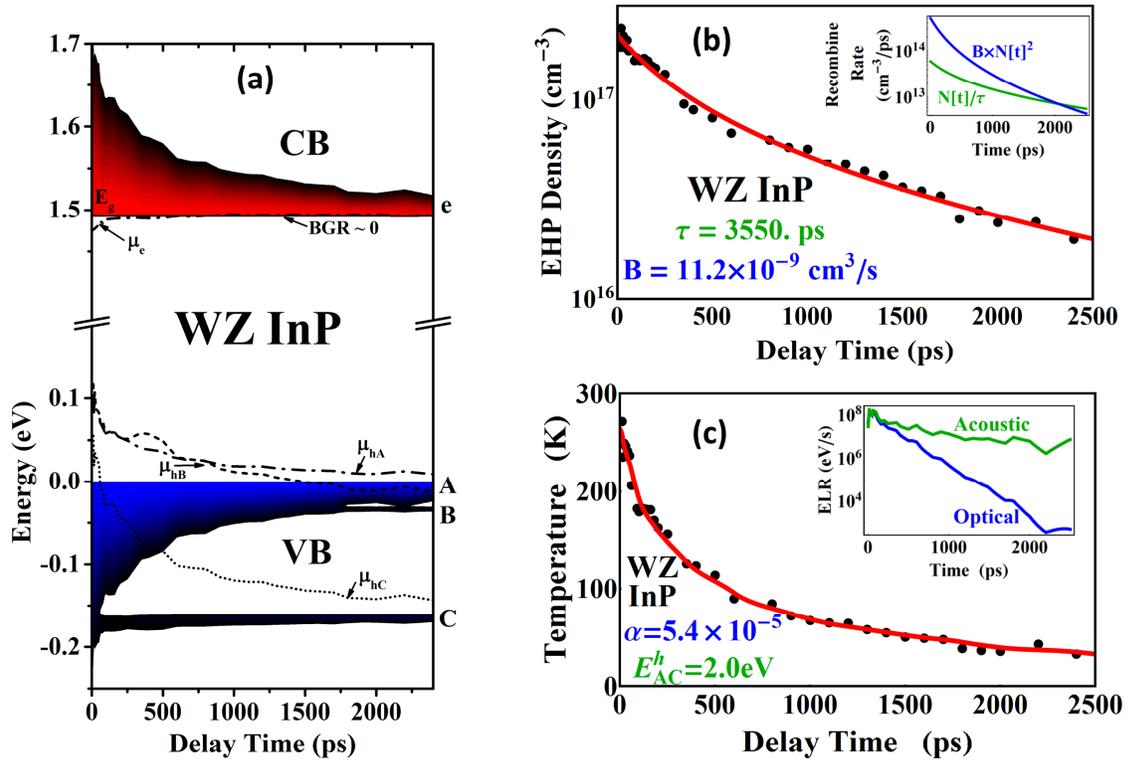

Fig.5 WZ InP nanowire TRS data fit results: (**a**) Electron-hole population density distributions, including quasi-Fermi-energies (dashed & dotted) and BGR (solid) in the conduction and valence bands as a function of time. (**b**) EHP density fit results (dots) as a function of time, with the red curve modeled by a sum of linear and bimolecular recombination processes. The inset shows the separate recombination rates. (**c**) EHP temperature fit results (dots) as a function of time, with the red curve modeled by optical and acoustic phonon scattering mechanism. The inset shows the energy loss rates due to optical or acoustic phonon emission.

With the dynamics of carrier density and temperature obtained from the fits to the TRS spectra, we can now plot the carrier population density distribution versus time for all conduction and valence bands. These are displayed in Fig.4a for ZB InP and Fig.5a for WZ InP. We assume that the background carrier density without pump excitation is $10^{15}$ cm$^{-3}$, so at late times when the carrier temperature is low, the occupation of the SO (for ZB) or B, C (for WZ) valence bands is still observable. For the same reason, the quasi-Fermi-energy for each band is computed individually from carrier density and temperature (with Fermi-Dirac distribution and 3D density of states) in spite of the assumption that pump excited holes in different hole bands are in thermal equilibrium, which means each hole band's quasi-Fermi-energies are equal. The fact that the quasi-Fermi-energy is usually within $k_B T$ of the band edge supports our assumption that the pump-excited carriers in the NWs are a degenerate EHP and thus we can ignore excitonic effects.

The EHP density and temperature obtained from these fits as a function of time are plotted in Fig.4b, c and 5b, c. The carrier density and temperature of the ZB InP NW is $1.5 \times 10^{18}$ cm$^{-3}$ and 400 K at t = 0, which relaxes to $7.5 \times 10^{16}$ cm$^{-3}$ and 70 K by 2 ns. In contrast, the carrier density and temperature of the WZ InP NW is $2 \times 10^{17}$ cm$^{-3}$ and 270 K at t = 0 which relaxes to $2 \times 10^{16}$ cm$^{-3}$ and 50 K by 2 ns. The difference of the initial carrier density and temperature reflects the fact that the pump excitation intensity for the ZB NW was $1.2 \times 10^{6}$ kW cm$^{-2}$ while for the WZ NW was lower at $4 \times 10^{5}$ kW cm$^{-2}$. The larger excitation intensity is also reflected by the fact that the ZB electron quasi-Fermi-energy is above the conduction band edge at the earliest times, while the WZ electron quasi-Fermi-energy is close to the conduction band edge. Similarly, the BGR of ZB InP NW is larger than WZ InP due to the higher EHP density.

To model the dynamics of the EHP carrier density, we include both linear (non-radiative) and bimolecular (radiative) recombination processes ignoring the background density from either unintentional doping or probe excitation. We describe the decay of the EHP density based on the sum of linear and bimolecular recombination by $\frac{dN}{dt} = -\frac{N}{\tau} - B \times N^2$ and obtain the non-radiative lifetime $\tau$ and bimolecular coefficient B for both ZB and WZ InP NWs. For the ZB InP NWs, $\tau \sim 1.4$ ns and B $\sim 2 \times 10^{-9}$ cm$^{-3}$ s$^{-1}$; while for the WZ InP NWs, $\tau \sim 3.4$ ns and B $\sim 11 \times 10^{-9}$ cm$^{-3}$ s$^{-1}$. These values compare reasonably well with previous reports [29-35, 45-46] taking into account that both parameters are strongly affected by sample quality, temperature and excitation carrier density. With the knowledge of $\tau$ and B, the density loss rate through both recombination mechanisms can be extracted. The values of the different loss rates as function of time are plotted in the insets of Fig.4b and 5b. Dividing the radiative recombination rate by the total recombination rate, we get the percentage of the carriers that recombine radiatively, or the internal quantum efficiency, as a function of time (see Fig. s2). The average value of quantum efficiency over 0 ~ 2 ns is 62% for ZB and 78% for the WZ InP NWs.

### Carrier thermalization dynamics: "Hot" phonon effects

The EHP temperature is related to the total carrier energy through the relation $E = \frac{3}{2} k_B T \frac{F_{\frac{3}{2}}(\eta)}{F_{\frac{1}{2}}(\eta)}$, where η is the quasi-Fermi-energy and $F_i(\eta)$ is the ith Fermi integral defined as usual. The decrease of carrier temperature is due to carrier energy loss. Usually the most significant energy loss mechanism for III-V semiconductors at higher carrier temperatures is phonon scattering with longitudinal optical (LO) phonon emission. At lower carrier temperatures, when the majority of carriers have energies below the LO phonon energy, acoustic-deformation-potential (ADP) scattering with acoustic phonon emission is dominant [36-38, 47-49]. Whether the optical or acoustic phonon emission dominates the energy loss rate (ELR) depends on *both* the carrier distribution and the phonon band structures, i.e. the LO, LA and TA phonon energies and density of states.

Examining the ZB InP NW energy loss rate (ELR) (displayed in Fig.4c inset), we find that the optical phonon ELR only slightly surpasses the acoustic phonon ELR at the earliest times, after which acoustic phonon ELR dominates. This contrasts with what is found in ZB GaAs [13] where the optical phonon ELR is more than an order of magnitude larger than the acoustic phonon ELR in the first 100 ps, after which acoustic phonon ELR dominants. Thus the overall carrier thermalization in the ZB InP NW is much slower than observed in the ZB GaAs NW, as is clearly observable in Fig. 6a. At late times, when acoustic phonon ELR dominates, the larger effective hole mass means that the relaxation will be dominated by the interaction of the acoustic phonons with holes [47-49] (See Fig. s3). We fit the temperature and find that the ZB InP NW hole acoustic deformation potential is 2.0 eV, which compares reasonably with previous reports considering its temperature dependence ( ~ 3 eV) [50-53].

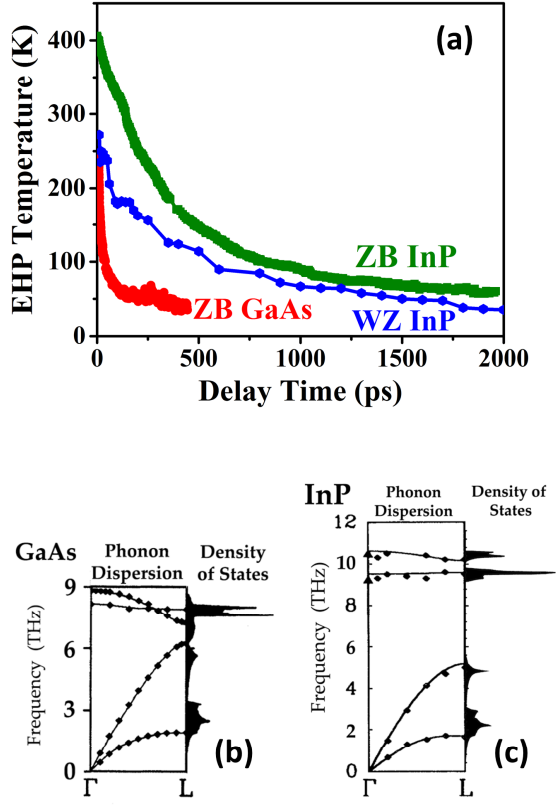

Fig. 6 (**a**) EHP temperature relaxation of ZB GaAs, ZB and WZ InP. The sharp contrast of early time temperature decay through LO phonon scattering is due to the distinction between (**b**) GaAs phonon band structure (Adapted from Giannozzi, P. et al [69]) and (**c**) InP phonon band structure (Adapted from Fritsch, J. et al [72]).

The distinctive difference between ZB GaAs and ZB InP NW optical phonon ELR can be modeled by an carrier-LO phonon Fröhlich coupling efficiency α introduced in the ELR equation $\langle \frac{dE}{dt} \rangle = \alpha \langle \frac{dE}{dt} \rangle_{LO} + \langle \frac{dE}{dt} \rangle_{ADP}$. As references [36-38, 47-49, 54-55] suggest, those LO phonon scattering ELR reduction are usually observed in transient measurements at highly excited (~$10^{18}$ cm$^{-3}$) semiconductors. Since this phenomenon does not exist in steady-state measurements or theoretical calculations based on Fröhlich interaction at the same temperature, previous reports [36, 54] attribute this phenomenon to the non-equilibrium or "hot" phonon effects. When the phonon temperature is elevated such that hot carriers' phonon emitting process begins to be compensated by the carriers' reabsorption of the hot phonons, the final LO phonon scattering ELR will be greatly reduced when the non-equilibrium phonon temperature approaches the carrier temperature.

The values of the Fröhlich coupling efficiency α for ZB GaAs and ZB InP NWs are found to be $\alpha^{ZB\ GaAs} = 3 \times 10^{-3}$ and $\alpha^{ZB\ InP} = 6 \times 10^{-5}$, respectively. The reason that hot phonon effects appear much stronger in ZB InP relative to ZB GaAs is that their phonon band structures are significantly different (see Fig. 6b, c), which implies a different pathway for LO phonon relaxation. After the initial pump pulse excitation, a non-equilibrium population of electrons and holes is relaxed initially and most rapidly by the emission of LO phonons. These LO phonons, in turn, usually decay into two acoustic phonons with equal energy and opposite momentum, e.g. LO → LA + LA, which is the case for ZB GaAs. In contrast, ZB InP has a large phonon band gap such that the LO(Γ) energy (10.5 THz) is more than twice the highest available acoustic phonon energy (4.8THz, LA at the L-points). Thus the ZB InP zone-center LO → LA + LA decay is not allowed. Instead, the ZB InP LO phonon must take a two-step process, i.e. LO → TO + TA / LA, then TO → LA(L) + LA(L), where the TO phonon energy (9.4 THz) is close to twice of the LA(L) phonon energy (4.8 THz). The ZB InP zone-center TO decay lifetime is reported [56-58] to be a factor of 1~2 longer than ZB GaAs zone-center LO phonon lifetime. The two-step processes suggest an even longer LO phonon relaxation for InP compared to the one-step process for GaAs LO phonon relaxation, which is consistent with the observation [58] that the ZB InP zone-center LO phonon lifetime is a factor of 3 ~ 4 longer than that of ZB GaAs at various temperatures. Thus the total ZB InP carrier energy loss through LO phonon emission per unit time is expected to be smaller than for ZB GaAs and the carrier cooling is expected to be slower. We believe that this qualitatively explains the fact that the initial temperature relaxation of the carriers though LO phonon scattering mechanism in ZB InP is almost 50-times less efficient than is observed in ZB GaAs. A detailed calculation of the LO phonon lifetime and Fröhlich coupling efficiency α is beyond the scope of this paper due to the complexity of the phonon emission and re-absorption dynamics.

Interestingly, for ZB InP NWs, we observe a signature at ~ 1.95 eV (Fig. 1a), which is about the energy of the valence band Γ-point to conduction band L-valley indirect transition [59-66]. We did not observe any signs of indirect transitions in ZB GaAs NW. In highly excited WZ InP NW, we only observe weak response from the direct transition between valence band and 2$^{nd}$ conduction band, which is zone-folded from the L-valley [67]. Hence the indirect transition should be less efficient and hard to observe. For ZB InP NWs, with the help of the LA(L) zone-edge phonons generated by hot zone-center LO phonon two-step decay process, the probe laser can relatively efficiently create indirect transitions thus causing a change in the scattering efficiency R' due to more absorption. Using the energy of the absorbed LA(L) phonon of 20 meV, we find the Γ-L indirect gap measured here is 1.97 eV, which is in the range of previous reports 1.94 ~ 2.2 eV at low temperatures [59-66].

Looking at the WZ InP NW ELR (Fig.5c inset), we find that the ratio of the initial ELR by LO and acoustic phonon scattering is similar to ZB InP NW, but still much less than that of ZB GaAs NWs. This is reflected by the value of the WZ InP NWs Fröhlich coupling efficiency $\alpha^{WZ\ InP} = 5.4 \times 10^{-5}$. The WZ InP hole acoustic-deformation-potential, fitted through a similar process as ZB InP, is found to be ~ 2 eV, which agrees with bulk ZB InP results [50-53].

While there is no direct report on WZ InP LO phonon decay lifetime either experimentally (recall that WZ InP does not exist in bulk form) or theoretically, we can still interpret this result using the same phenomenological decay channel framework as detailed above for ZB InP. For instance, the WZ InP phonon band structures at Γ-A dispersion along [0001] can be roughly approximated [59] by folding the ZB InP Γ-L of the phonon band structure along [111]. The phonon states are now distributed closer to the zone-center, but the top of the acoustic branches should be still at exactly the same energy as ZB InP and the dispersion of LO/TO branches would not change due to the flat dispersion nature of the optical branches. Hence, the direct LO → LA + LA process is still not allowed for WZ InP and a similar two-step process like that of ZB InP is required. For that reason, we would expect a similar hot phonon effect for WZ InP NW and thus a similarly slow thermalization process as in ZB InP NW, in agreement with our observations.

## Conclusions

Using time-resolved Rayleigh scattering we have directly measured the density and temperature relaxation of optically excited electron-hole plasma in single ZB and WZ InP nanowires. We demonstrate that a simple analysis technique enables the extraction of the electron and hole density and temperature in complex multi-band materials. Compared with similar measurements of ZB GaAs nanowires, we find that the thermalization time of the EHP is nearly 50 times longer in ZB InP nanowires. This slower thermalization is likely caused by the substantially different phonon band structure of InP compared to GaAs which inhibits the direct relaxation of optical to acoustic phonons. Both ZB InP nanowires and WZ InP nanowires have similar thermalization dynamics. Information on the dependence of carrier thermalization dynamics in different materials and nanostructures is an essential component for the design of high efficiency nanowire devices.

AUTHOR INFORMATION

**Corresponding Author**

*E-mail address: leigh.smith@uc.edu

ACKNOWLEDGMENTS

We acknowledge the financial support of the National Science Foundation through grants DMR-1105362, 1105121, and ECCS-1100489 and the Australian Research Council. The Australian National Fabrication Facility is acknowledged for access to the growth facilities used in this research.